\documentstyle[prl,aps,twocolumn]{revtex}
\large  

\input epsf
\begin{document}
\draft \twocolumn[\hsize\textwidth
\columnwidth\hsize\csname
@twocolumnfalse\endcsname
\title{Low-dimensional Bose liquids: beyond the Gross-Pitaevskii
approximation}
\author{Eugene B. Kolomeisky $^{(1)}$, T. J. Newman $^{(1)}$, 
Joseph P. Straley$^{(2)}$, and Xiaoya Qi$^{(1)}$} 
\address{$^{(1)}$ Department of Physics, University of Virginia,
McCormick Road, Charlottesville, VA 22903\\
$^{(2)}$ Department of Physics and Astronomy, University of Kentucky,
Lexington, KY 40506}
\maketitle
\begin{abstract}
The Gross-Pitaevskii approximation is a long-wavelength theory widely used to
describe a variety of properties of dilute Bose condensates, in particular
trapped alkali gases. We point out that for short-ranged repulsive
interactions this theory fails in dimensions $d \le 2$, and we propose the
appropriate low-dimensional modifications. For $d=1$ we analyze density
profiles in confining potentials, superfluid properties, solitons,
and self-similar solutions.
\end{abstract}
\vspace{2mm} 
]

\narrowtext

Experimental observation of Bose-Einstein condensation (BEC) in trapped
alkali vapors \cite{ander} has ushered in a new era of superlow temperature
physics. The Gross-Pitaevskii (GP) mean-field theory \cite{gp} has proven
to be an indispensable tool both in analyzing and predicting the outcome
of experiments.

With rapid progress in the experimental exploration of BEC systems it is
reasonable to anticipate that effectively one and two-dimensional systems
are realistic prospects in the near future \cite{jack}. For example, high
aspect ratio cigar-shaped traps approximating quasi-one-dimensional BEC
systems are already available experimentally \cite{andr}. From the 
theoretical viewpoint low-dimensional systems are rarely well represented
by mean-field theories which leads one to question the validity of 
the GP theory in one and two dimensions. In this paper we shall show
that, indeed, the physics of dilute Bose systems requires a 
fundamental modification of the GP theory in low dimensions $d \le 2$.

The GP theory is a quasi-classical (or mean-field) approximation which
replaces the bosonic field operator $\psi$ by a classical order parameter field
$\Phi ({\bf r},t)$. For short-ranged interactions 
the interparticle potential $U({\bf r})$ is replaced by $g\delta ^{d}
({\bf r})$, where $g$ is the pseudopotential. Then for a system of 
bosons in an external potential $V$, 
the energy functional and the equation of motion for $\Phi$
take the following form:
\begin{equation}
\label{gp1}
F_{GP}=\int d^{d}r \left [ {\hbar ^{2}\over 2m} |\nabla \Phi|^{2}
+ V({\bf r})|\Phi |^{2} + {g \over 2}|\Phi |^{4} \right ] \ ,
\end{equation}
and
\begin{equation}
\label{gp2}
i \hbar \partial _{t} \Phi = {\delta F_{GP}\over \delta \Phi ^{*}}
= \left [ -{\hbar ^{2}\over 2m} \nabla ^{2} 
+ V({\bf r}) + g|\Phi |^{2} \right ] \Phi \ .
\end{equation}
The GP equations (\ref{gp1}) and (\ref{gp2}) are widely used to compute a 
variety of properties of Bose systems \cite{gp}.

The GP approximation is a long-wavelength theory relying on the concept of
the pseudopotential to account for interparticle interactions. However, for
repulsive bosons the
canonical pseudopotential vanishes in two dimensions \cite{schick}, 
implying that an essential modification of the GP theory is necessary for
$d\le 2$. To see how to modify the theory, it is useful to rewrite the
last integrand of (\ref{gp1}) in terms of the particle density
$n = |\Phi|^{2}$, so that $(g/2)|\Phi |^{4} = (g/2)n^{2}$. This can be
recognized \cite{gp,kol1} as the lowest order term of a dilute expansion
of the ground-state energy density for $d>2$. Thus the correct low-dimensional
local density theory will instead
have the ground-state energy density of the $d \le 2$
dilute Bose system\cite{schick,kol1}, which is not proportional to $n^{2}$. 

Let us write the interparticle interaction in the form $U({\bf r}) = 
u_{0}\delta _{a}^{d}({\bf r})$ where $u_{0}$ is the amplitude of the
interparticle repulsion, and the notation $\delta _{a}^{d}({\bf r})$ denotes
any well-localized function that transforms into the mathematical
Dirac $\delta$ function when the range of interaction $a \rightarrow 0$.
The renormalization group analysis of this problem \cite{kol1} reveals
that two dimensionless combinations $na^{d}$ and $\hbar ^{2}n^{(2-d)/d}/
mu_{0}$ play an important role in defining the conditions of the dilute
limit for $d \le 2$. 

In the dilute limit $na \ll 1$ and $\hbar ^{2}n/mu_{0}\ll 1$,
{\it any} one-dimensional Bose system with short-ranged
repulsive interactions becomes equivalent to a gas of free-fermions
\cite{gl} (or equivalently point hard-core
bosons\cite{note1}), with energy density  $\pi^{2}\hbar ^{2}n^{3}/6m$.
This can be generalized\cite{kol1} for $d < 2$ to the statement that
for  $na^{d} \ll 1$ and $\hbar ^{2}n^{(2-d)/d}/mu_{0} \ll 1$,
the lowest order term of the ground-state energy density expansion
is {\it universal} and given by  $(\hbar ^{2}C_{d}/2m)n^{(2+d)/d}$,
where $C_{d}$ is a $d$-dependent constant.   This implies that the quartic
nonlinearity $|\Phi |^{4}$ in Eq.(\ref{gp1}) should be replaced by
$|\Phi |^{2(2+d)/d}$.   
In particular, for the  practically important case of
one dimension, the system of equations (\ref{gp1}) and (\ref{gp2})
is modified to
\begin{equation}
\label{newgp1}
F={\hbar ^{2}\over 2m}\int dx \left [ \left | {d\Phi\over dx} \right | ^{2}
+ {2m\over \hbar ^{2}} V(x)|\Phi |^{2} + {\pi ^{2} \over 3}
|\Phi |^{6} \right ] \ ,
\end{equation}
and
\begin{equation}
\label{newgp2}
i \hbar \partial _{t} \Phi  
= {\hbar ^{2} \over 2m} \left [ -\partial _{x}^{2} 
+ {2m \over \hbar ^{2}} V(x) + \pi ^{2}|\Phi |^{4} \right ] \Phi \ .
\end{equation}

Similar considerations applied to the marginal two-dimensional case lead
to the conclusion that in the dilute limit \cite{schick,kol1} a theory
replacing the GP approximation starts from the energy functional
\begin{equation}
\label{newgp2d}
F={\hbar ^{2}\over 2m}\int d^{2}r \left [ |\nabla \Phi|^{2}
+ {2m\over \hbar ^{2}} V(x)|\Phi |^{2} + {4\pi ^{2}\over
|\ln (|\Phi|^{2}a^{2})|}|\Phi |^{4} \right ] \ .
\end{equation}
Ignoring the logarithmic factor will perhaps suffice for many
practical purposes; then (\ref{newgp2d}) is precisely of the GP form.
Finding an experimental manifestation of the logarithmic correction
is likely to be very challenging.

A detailed analytical and numerical study of the one and two-dimensional
cases will be given in a longer publication\cite{knsq}; hereafter we 
restrict ourselves to only the salient features of one dimension where
the deviations from the GP theory are largest.

\vspace{0.1cm}

\noindent
{\bf Density profiles in external potentials:}
The stationary solution to Eq.(\ref{newgp2}) defined via 
$\Phi (x,t) = \phi (x)e^{-i\mu t/\hbar}$ can be found by solving 
\begin{equation}
\label{stat1}
{d^{2}\phi \over dx^{2}} + {2m\over \hbar^{2}}(\mu - V(x))\phi
-\pi^{2}\phi ^{5} = 0 \ ,
\end{equation}
subject to the condition of fixed total particle number $N=\int dx \phi^{2}$
which determines the chemical potential $\mu $. For an external potential
that varies slowly on the scale of the interparticle spacing the derivative
term in (\ref{stat1}) can be ignored: this gives the density profile in
the Thomas-Fermi (TF) approximation:
\begin{equation}
\label{tfa}
n_{TF}(x) = \phi _{TF}^{2} = \left [ 2m(\mu - V(x)) \right ]^{1/2}/
\pi \hbar \ ,
\end{equation}
with the density being zero in the classically forbidden region $\mu < V(x)$.
For the practically important case of a harmonic trap, $V=m\omega^{2}x^{2}/2$,
and the density profile is elliptical:
\begin{equation}
\label{tfa1}
n_{TF}(x) = \left [ (2m\mu - m^{2}\omega^{2}x^{2}) \right ]^{1/2}/
\pi \hbar \ .
\end{equation}
The chemical potential is given by $\mu _{TF} = \hbar \omega N$, the density
in the center of the trap is $n_{TF}(0) = (2m\hbar \omega N)^{1/2}/\pi \hbar$,
and the size of the trapped condensate is $2(2m\hbar \omega N)^{1/2}/
\pi \hbar$.

The accuracy of these predictions can be tested against the exact solution
of a dilute system of bosons with repulsive interactions: an ideal
candidate being a system of point impenetrable bosons. The boson-fermion
equivalence\cite{gl} implies that in the many-body system, the single-particle
energy levels $E_{n}=\hbar \omega (n+1/2)$ of the harmonic oscillator are
occupied in a fermionic fashion, i.e. with no more than one particle per state.
The chemical potential is then given by $\mu = \hbar \omega (N+1/2)$, which 
for large $N$ approaches our TF result $\mu _{TF} = \hbar \omega N$.
Similarly the density profile can be computed as a sum of squares of the
single-particle wave functions:
\begin{equation}
\label{ba}
n(x) = {1\over (\pi l)^{1/2}} \sum \limits _{k=1}^{N-1}
{1 \over 2^{k}k!}H_{k}^{2}(x/l)\exp (-x^{2}/l^{2}) \ ,
\end{equation}
where $H_{k}$ are Hermite polynomials and $l=(\hbar /m\omega)^{1/2}$.
The density distribution (\ref{ba}) is plotted in Fig.1 where it is 
compared with a) the numerical solution of (\ref{stat1}) 
with $V=m\omega^{2}x^{2}/2$, and b) the TF result (\ref{tfa1}), 
for different numbers of particles.
The main flaw of the theory based on 
(\ref{stat1}) is that it does not reproduce density oscillations due to
algebraic ordering of the particles. This is not surprising as (akin to 
the GP approximation) the discreteness of the particles, which is 
responsible for the density oscillations, is ignored. Otherwise, the agreement
between the approximate and the exact profiles is very good; in the limit
of large particle number the differences become imperceptible. 
These results can be directly tested experimentally; as a 
comparison we note that the one-dimensional GP theory in the TF approximation
predicts \cite{gp} $n_{TF} \sim \mu - V(x)$, which is quite distinct 
from (\ref{tfa}), and agrees very poorly with the exact result.

\begin{figure}[htbp]
\epsfxsize=3.0in
\vspace*{-0.3cm}
%\hspace*{0.2cm}
\epsfbox{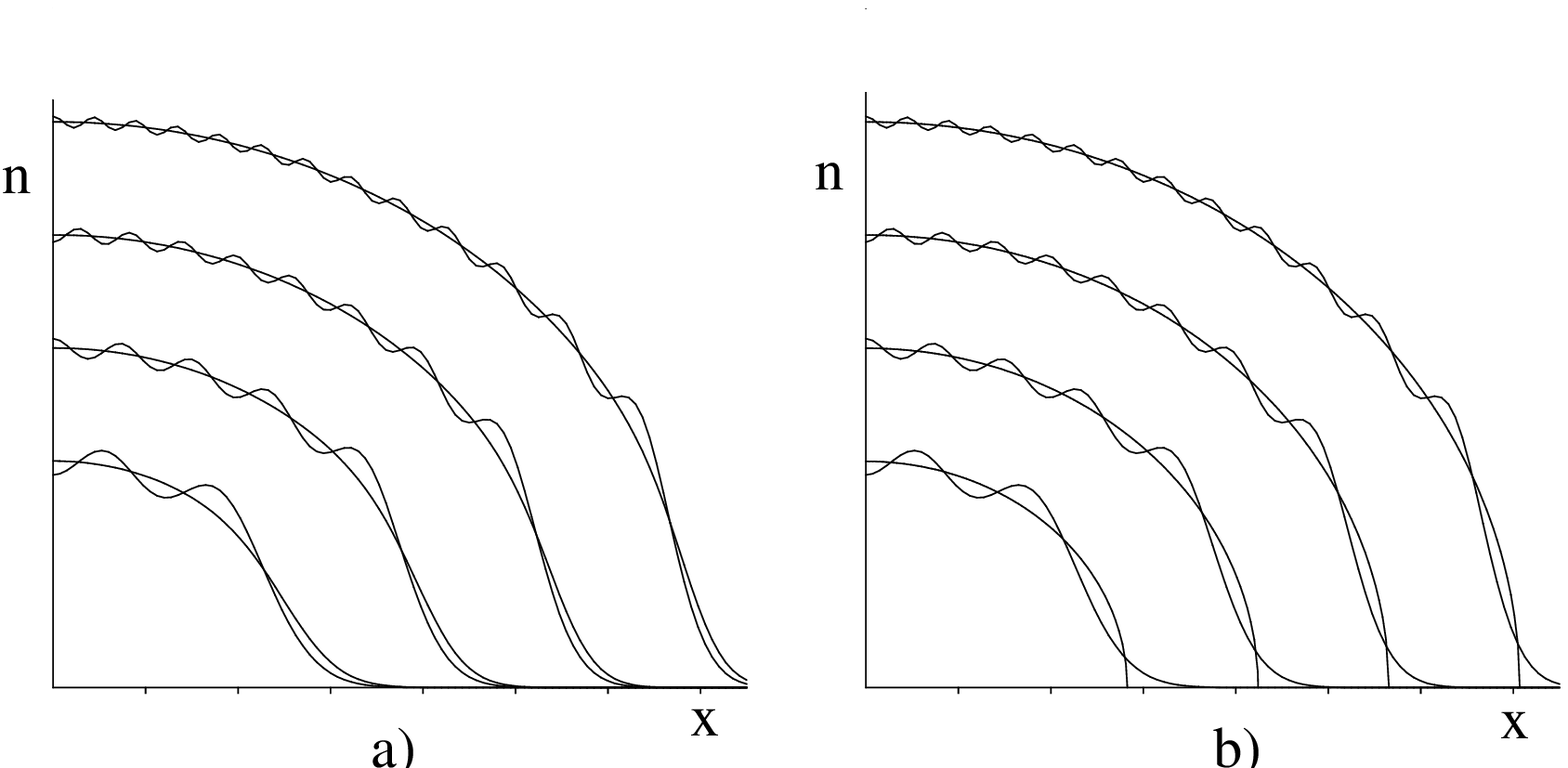}
\vspace*{0.1cm}
\caption{The density profile (\ref{ba}) plotted for particle numbers
$N=4,9,16,25$ in units $m\omega /\hbar = 1$.
The non-oscillating curves correspond to a) the numerical solution 
of (\ref{stat1}) and b) the TF result (\ref{tfa1}).}
\end{figure} 

\noindent
{\bf Solitons:} Gray solitons \cite{kl} have been recently created and their
dynamics was observed in cigar-shaped condensates of $^{87}Rb$ vapors 
\cite{burg}, which makes it important to understand solitonic properties
of the system (\ref{newgp1}) and (\ref{newgp2}). Let us look for solutions
to (\ref{newgp2}) (with $V=0$) of the form $\Phi (x,t) = \phi (x,t)
e^{-i\mu t/\hbar} $. The function $\phi$ then obeys the equation
\begin{equation}
\label{sol1}
i\hbar \partial _{t}\phi = {\hbar ^{2}\over 2m} \left [ -\partial _{x}^{2}
\phi + \pi ^{2} (|\phi|^{4} - \phi_{0}^{4})\phi \right ] \ ,
\end{equation}
where the chemical potential $\mu = \pi ^{2}\hbar ^{2}\phi _{0}^{4}/2m$
is selected so that the particle density $n_{0}=\phi _{0}^{2}$ is constant
at infinity. In dimensionless variables $f=\phi/\phi _{0}$, $y=\pi n_{0}x$,
$\tau = \pi^{2}n_{0}^{2}\hbar t/m$, Eq.(\ref{sol1}) simplifies to
\begin{equation}
\label{sol2}
2 i \partial _{\tau}f = -\partial _{y}^{2}f
 + (|f|^{4} - 1)f  \ .
\end{equation}

We will be looking for a localized traveling wave solution \cite{note2}
to (\ref{sol2}) of the form $f(y,\tau) = f(y-\beta \tau)$ where the
dimensionless velocity $\beta $ is measured in units of the sound velocity
$c=\pi \hbar n_{0}/m$. This problem can be solved exactly. The results
are conveniently described in terms of the amplitude $A$ and phase $\theta$
of the dimensionless order parameter $f = Ae^{i\theta}$:
\begin{eqnarray}
\label{solres}
\nonumber
A^{2} & = & 1 - {3(1-\beta ^{2}) \over 2 + (1+3\beta ^{2})^{1/2} 
{\rm cosh}[2(1-\beta ^{2})^{1/2}(y-\beta \tau)] } \\
2\theta & = & {\rm cos}^{-1} \left [ {(3\beta ^{2}/A^{2}) - 1 \over
(1+3\beta ^{2})^{1/2} } \right ] \ .
\end{eqnarray}
The spatial behavior given by (\ref{solres}) is shown in Fig.2.

\begin{figure}[htbp]
\epsfxsize=3.0in
\vspace*{-0.1cm}
%\hspace*{0.2cm}
\epsfbox{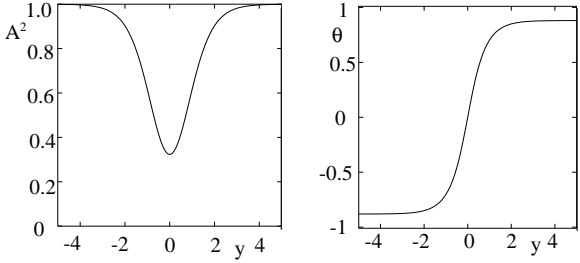}
\vspace*{0.1cm}
\caption{The density $A^{2}$ and the phase $\theta$, 
for the moving soliton (\ref{solres}) with $\beta=1/2$.}
\end{figure} 

The amplitude in (\ref{solres}) describes a moving depression (particle
deficit) with the minimal value at the soliton center given by
$A^{2}(0) = (1+3\beta ^{2})^{1/2}-1$. The soliton exists only for
$\beta < 1$ (i.e. the soliton velocity cannot exceed the speed of sound);
for $\beta = 1$ (\ref{solres}) gives the uniform result $A^{2}=1$. On the
other hand, for $\beta = 0$ (i.e. a vortex, or dark soliton \cite{kl})
the minimal value of the amplitude at the soliton center drops to zero.

The phase expressed in (\ref{solres}) varies rapidly in the vicinity of
the amplitude dip, staying approximately constant far away from it. The
total phase shift across the soliton can be found as 
$\Delta \theta = {\rm cos}^{-1}[(3\beta ^{2}-1)/(1+3\beta ^{2})^{1/2}]$. It is 
a continuous function of the soliton velocity varying between $\pi $ (when
$\beta =0$) and zero (when $\beta = 1$). Antisolitons may be defined as having
opposite signs of $d\theta /dy$, and there are no constraints on $\Delta
\theta $ for the open line or ring geometries (provided the number of
solitons matches the number of antisolitons). However, if there is an 
imbalance of solitons and antisolitons in the ring geometry, then the
uniqueness of the order parameter $f(y,\tau)$ implies that $\Delta \theta $
is a fraction of $2\pi $ for any excess soliton; this will in turn mean that
the excess soliton velocity is {\it quantized}.

The solution (\ref{solres}) bears some similarity with the one-dimensional
soliton of the GP theory \cite{kl}; the main qualitative difference
(seen after recovering the original units) is that in the dilute limit
the soliton size is of order $1/(1-\beta ^{2})^{1/2}n_{0}$ {\it independent}
of the amplitude of the interparticle repulsion \cite{note3}.

General methods \cite{kl} can be used to compute the soliton energy $E$ and
momentum $P$. For their dimensionless counterparts $\epsilon = 2mE/
\pi^{2}\hbar ^{2}n_{0}^{2}$ and ${\cal P}=P/\pi \hbar n_{0}$ we find
\begin{eqnarray}
\label{enmom}
\nonumber
\epsilon & = & {\surd{3}\over \pi} (1-\beta ^{2}) \ln \left \lbrace
{2 + [3(1-\beta ^{2})]^{1/2} \over (1+3\beta ^{2})^{1/2}} \right
\rbrace \\
{\cal P} & = & -{\beta \over (1-\beta ^{2})} \epsilon
+ {1 \over \pi} {\rm cos}^{-1} \left [ {3\beta ^{2}-1 \over
(1+3\beta ^{2})^{1/2}} \right ] \ . 
\end{eqnarray}

The dependencies $\epsilon (\beta )$ and ${\cal P}(\beta )$ parametrically
define the soliton dispersion law $\epsilon ({\cal P})$ which should
be identified \cite{note4} with the ``hole'' branch of the elementary
excitations spectrum \cite{lieb}. To assess the accuracy of 
$\epsilon ({\cal P})$ given in (\ref{enmom}) we compare it with the exact 
result of Lieb\cite{lieb} for the system of $\delta$-interacting bosons
in the dilute limit $\hbar ^{2}n/mu_{0} \ll 1$: $\epsilon _{\rm exact}
({\cal P}) = 2|{\cal P}| - {\cal P}^{2}$, for $|{\cal P}| \le 1$.
Since the velocity $\beta $ in (\ref{enmom}) varies between zero and
unity, the momentum (which we choose to be positive) computed from 
(\ref{enmom}) varies between unity and zero in correspondence with the
exact result. It is straightforward to show that for ${\cal P} \ll 1$, 
the elimination of $\beta $ in (\ref{enmom}) leads to $\epsilon = 2{\cal P}$
which is again in agreement with the exact result. The behavior 
$\epsilon ({\cal P})$ implied by (\ref{enmom}) in the vicinity of the
end-point of the spectrum ${\cal P}=1$ is qualitatively similar, and
quantitatively close to the exact dependence. To illustrate these statements
we have plotted the dispersion law (\ref{enmom}) in Fig.3 against the
exact result.

\begin{figure}[htbp]
\epsfxsize=3.0in
\vspace*{-0.1cm}
%\hspace*{0.2cm}
\epsfbox{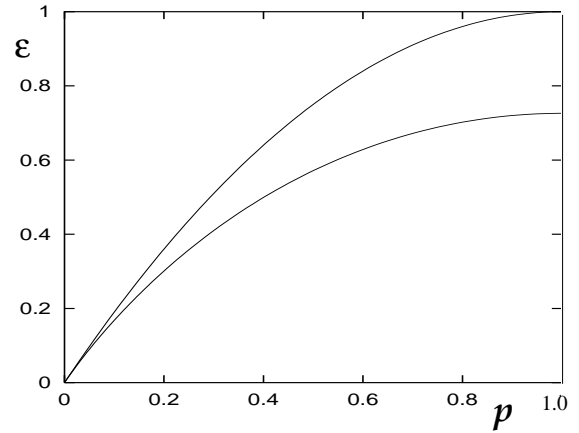}
\vspace*{0.1cm}
\caption{The spectrum parameterized by (\ref{enmom}) (lower curve)
compared to the exact result of Lieb (upper curve).}
\end{figure} 
 
\noindent
{\bf Superflow:} The dimensionless current density is given by $j=A^{2}\partial
_{y}\theta$, and below we look for solutions with fixed given current
$j$ (i.e. the steady state) and $\partial _\tau A = \partial _\tau \theta
= 0$. Substituting $f=Ae^{i\theta}$ and $j=A^{2}d\theta/dy$ into (\ref{sol2}),
and imposing fixed chemical potential, we find:
\begin{equation}
\label{sfamp}
{d^{2}A\over dy^{2}} = {j^{2}\over A^{3}} + A^{5} - A \ .
\end{equation}
In the spatially uniform state ${d^{2}A\over dy^{2}}=0$ and one finds the 
dimensionless amplitude
$A_{\infty}^{4} = [1 + (1-4j^{2})^{1/2}]/2$,
which implies that superflow {\it reduces} the amplitude of the 
order parameter. The uniform solution, and thus superfluidity,
cease to exist above the critical flow $j_{c} = 1/2$ when the amplitude
drops down to its minimal value $A_{\infty}^{c}=2^{-1/4}$. These results
imply that the critical velocity for superfluidity in the original units
is $c/\surd{2}$.

The equation (\ref{sfamp}) also has an immobile well-localized solution
in the form of a dip of the order parameter; far away from the dip the 
amplitude recovers to its uniform value. The dip solution
is closely related to the soliton previously discussed. Indeed,
in the reference frame moving with the flow, the dip solution is moving
and thus is identical to a soliton. The functional form of the
dip can be deduced from (\ref{solres}) by replacing $\beta$ by
$j/A_{\infty}^{4}$, $A$ by $A/A_{\infty}$, and $(y-\beta \tau)$ by
$yA_{\infty}^{2}$. The dip solution disappears altogether for $j>j_{c}$.

\vspace{0.1cm}

\noindent
{\bf Self-similar solutions:} The results derived so far have their
counterparts in the context of the one-dimensional GP approximation. However
the theory based on Eqs.(\ref{newgp1}) and (\ref{newgp2}) allows
self-similar solutions which do not exist in the one-dimensional GP
theory \cite{note5}. Below we only look at the cases consistent with 
the condition of conservation of total particle number. Consider the 
system of bosons placed in a harmonic trap. In dimensionless variables
it is described by
\begin{equation}
\label{harmeom}
2i\partial _{\tau}f = -\partial _{y}^{2}f + [|f|^{4}
+ w^{2}y^{2}]f \ ,
\end{equation}
where $w=m\omega/\pi^{2}\hbar n_{0}^{2}$ is the dimensionless oscillator
frequency ($n_{0}$ now has the meaning of a density introduced to make
$f$ dimensionless; it should be determined from the complete solution
of the problem). In contrast to (\ref{sol2}) (and without loss of generality)
we have shifted the origin of the chemical potential. The self-similar
solution $f=Ae^{i\theta}$ derived from (\ref{harmeom}) has the form
\cite{note5}
\begin{equation}
\label{sss}
A=\rho(\tau)^{-1/2}h(y/\rho(\tau)) \ , \ \ \ \theta = \theta _{0}(\tau)
+ {1\over 2} {d\ln \rho \over d\tau} y^{2} \ ,
\end{equation} 
where the functions $\rho (\tau)$ and $h(v)$ obey the equations
\begin{eqnarray}
\label{auxeqsa}
{d^{2}\rho \over d\tau ^{2}} & = & -w^{2}\rho + {\gamma \over \rho ^{3}} \\ 
\label{auxeqsb}
{d^{2}h \over dv^{2}} & = & -\delta h + h^{5} + \gamma v^{2}h 
\end{eqnarray}
where $\gamma$ and $\delta$ are arbitrary constants. Eq.(\ref{auxeqsb}) for the
scaling function $h(v)$ has localized solutions only for $\delta > 0$
and $\gamma \ge 0$: for $\gamma = 0$ an explicit analytic solution to
(\ref{auxeqsb}) can be written down, while for $\gamma > 0$, (\ref{auxeqsb})
has the same functional form as the equation we encountered in determining
the density profile in the harmonic trap (cf. (\ref{stat1}) for $V=m\omega^{2}
x^{2}/2$].

The dynamics of the length scale $\rho (\tau)$ can be understood by viewing
(\ref{auxeqsa}) as a fictitious classical mechanics problem in the 
potential $U=w^{2}\rho ^{2}/2 + \gamma/2\rho^{2}$. This analogy implies
that an initially localized cloud of bosons in free space
($w=0$) will expand asymptotically 
in a ballistic fashion: $\rho (\tau) \sim \tau$. 
In the presence of the confining potential ($w \ne 0$) the scale
$\rho (\tau)$ oscillates between maximum and minimum values: for 
$\gamma=0$ the dynamics of $\rho $ is the same as for a harmonic
oscillator of frequency $w$.

\vspace{0.2cm}

We have also performed direct numerical integration of the non-linear
equation (\ref{newgp2}), and have confirmed the existence of both the 
similarity solutions, and moving trains of solitons with quantized 
velocity, with amplitude and phase as given by (\ref{solres}).
More details will be given in a future publication\cite{knsq}.

In conclusion we have presented a new continuum description of 
dilute Bose liquids
appropriate for low dimensional systems. 
For the case of one dimension, we have derived stationary
properties, along with solitonic and similarity solutions. In particular,
the latter have no analog in the GP theory. It is our hope that these
results will be testable in BEC experiments in the near future.

\end{document}